\documentstyle[aps,amssymb]{revtex}

\newcommand{\bra}[1]{\mbox{$\langle \, {#1}\, |$}}
\newcommand{\ket}[1]{\mbox{$| \, {#1}\, \rangle$}}
\newcommand{\exval}[1]{\mbox{$\langle \, {#1}\, \rangle$}}
\newcommand{\be}{\begin{equation}}
\newcommand{\bel}[1]{\begin{equation}\label{#1}}
\newcommand{\ee}{\end{equation}}
\newcommand{\bea}{\begin{eqnarray}}
\newcommand{\ba}{\begin{array}}
\newcommand{\eea}{\end{eqnarray}}
\newcommand{\ea}{\end{array}}

\begin{document}
\draft
\tighten
\onecolumn
\twocolumn[\hsize\textwidth\columnwidth\hsize\csname @twocolumnfalse\endcsname

\title{Branching-annihilating random walks in one dimension:
Some exact results}
\vspace {1truecm}

\author{K. Mussawisade$^1$, J. E. Santos$^2$ and G. M. Sch\"utz$^1$}

\vskip 0.5truecm

\address{{}$^1$Institut f\"ur Festk\"orperforschung,
Forschungszentrum J\"ulich, 52425 J\"ulich, Germany}

\address{{}$^2$Department of Theoretical Physics,
University of Oxford, 1 Keble Road, Oxford, OX1 3NP, United Kingdom}

\vskip 0.5truecm

\date{\today}
\vskip 1truecm
\maketitle

\begin{abstract}
{
We derive a self-duality relation for a one-dimensional model of
branching and annihilating random walkers with an even number of
offsprings. With the duality relation and by deriving exact results
in some limiting cases involving fast diffusion we obtain new information
on the location and nature of the phase transition line between an
active stationary state (non-zero density) and an
absorbing state (extinction of all particles), thus clarifying
some so far open problems. In these limits the transition is mean-field-like,
but on the active side of the phase transition line the fluctuation in the
number of particles deviates from its mean-field value. We also show
that well within the active region of the phase diagram a {\em finite}
system approaches the absorbing state very slowly
on a time scale which diverges exponentially in system size. \\[2mm]
}~
\end{abstract}

\vskip2pc]

\section{Introduction}
\label{S1}
In a branching-annihilating random walk (BARW) particles hop on a
lattice, annihilate pairwise on encounter, but may also spontaneously
create offsprings on the same or on nearest neighbour lattice sites.
Such models appear in a large variety of contexts, in particular in
reaction-diffusion mechanisms and in non-equilibrium spin
relaxation. A generic feature of these processes is a transition
as a function of the annihilation and branching rates between an active
stationary state with non-zero particle density and an absorbing,
inactive state in which all particles are extinct. Numerical results
gained from a large variety of systems suggest that the transition in
models with a single (or an odd number) of offsprings fall into the
universality class of directed percolation (DP) \cite{DP}, whereas models
with an even number of offsprings belong to a distinct parity-conserving
(PC) universality class \cite{PC,Meny94}. A coherent picture of this
scenario is provided from an renormalization point of view \cite{Card97}.

In this paper we use exact methods to derive a self-duality relation
and to address some open questions for limiting cases of the
BARW model of Ref. \cite{Meny94} which is a model for spin relaxation
dynamics far from thermal equilibrium. One considers Ising spins
in one dimension with generalized zero-temperature Glauber dynamics
\cite{Glau63}, but with an independent coupling to an infinite-temperature
heat bath which allows for Kawasaki spin-exchange events \cite{Kawa66}
with rate $\alpha/2$. This spin-flip process can be visualized in the
following way:
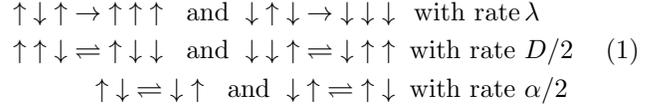
\begin{eqnarray}
\uparrow \; \downarrow \; \uparrow \; \to \; \uparrow \; \uparrow \;
\uparrow \; \mbox{ and }
\downarrow \; \uparrow \; \downarrow \; \to
 \; \downarrow \;  \downarrow \;
\downarrow \; & \mbox{with rate} & \lambda \nonumber \\
\label{1-1}
\uparrow \; \uparrow \; \downarrow \;
\rightleftharpoons \;
\uparrow \; \downarrow \; \downarrow \;
\mbox{ and }
\downarrow \; \downarrow \; \uparrow \;
\rightleftharpoons \; \downarrow \;  \uparrow \;
\uparrow \; & \mbox{with rate } & D/2  \\
\uparrow \; \downarrow \; \rightleftharpoons \; \downarrow \; \uparrow \;
\mbox{ and }
\downarrow \; \uparrow \; \rightleftharpoons \; \uparrow \;
\downarrow \; & \mbox{with rate } & \alpha/2 \nonumber
\end{eqnarray}
By identifying a domain wall ($\uparrow\;\downarrow$ or
$\downarrow\;\uparrow$) with a particle of type $A$ on the dual lattice
and two parallel spins with
a vacancy $\emptyset$ \cite{Racz85} this process becomes a BARW with rates

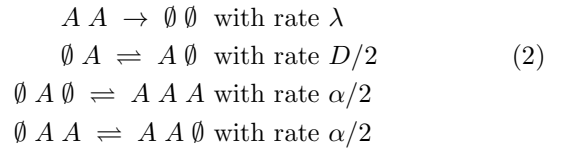
\begin{eqnarray}
 A \; A \;\to \;\emptyset\;\emptyset \;
& \mbox{ with rate } & \lambda \nonumber \\
\label{1-2}
\emptyset \; A \; \rightleftharpoons
 \; A\;\emptyset\; & \mbox{ with rate } & D/2  \\
\emptyset \; A \; \emptyset \; \rightleftharpoons \; A \; A \; A
& \mbox{ with rate } & \alpha/2 \nonumber \\
\emptyset \; A \; A \; \rightleftharpoons \; A \; A \; \emptyset \nonumber
& \mbox{ with rate } & \alpha/2 \nonumber
\end{eqnarray}
Without branching (i.e. spin exchange) the
system evolves into the single absorbing state with no particles at all.
In spin language this is the totally ferromagnetic state with all spins
up or all spins down. In the presence of the branching process an
intricate competition between the zero-temperature ordering process
(particle annihilation) and the disordering high-temperature branching
process sets in. The result is a non-trivial phase diagram as a function
of the system parameters.
Starting from, say, a random initial state with an even number of
particles the system evolves ultimately into the inactive empty lattice
for dominant ordering dynamics, whereas it remains in an active state
with finite density if the disordering branching process dominates.
Numerical evidence suggests the phase transition to belong to the
PC universality class \cite{Meny94}.

We stress that these results are supposed to be valid only
in the thermodynamic limit. In any {\em finite} system the unique
stationary state is the absorbing inactive state (unless $\lambda=0$)
because for $\lambda > 0$ there is always a small probability of reaching
this state from which the system cannot escape any more. However,
intuitively, one expects the approach to this state to occur on a time
scale $\tau_{act} \sim q^L$ which is exponentially large in system
size $L$ if parameters are chosen to represent the active phase of the
thermodynamic limit. In the absorbing phase, both exact analytical results
for $\lambda=D$ \cite{Droz89} and renormalization group results on
diffusion-limited annihilation ($\alpha=0$) \cite{Lee94} show that the
approach to extinction is algebraic for the infinite system. For a finite
system one can infer from these results a crossover time scale
$\tau_{abs} \sim L^2$ to exponential decay of the particle density.

Here we aim at obtaining information on the form of the phase transition
line and on the dynamical and stationary behaviour of the system in
various limiting cases involving fast diffusion of particles (or spins
in terms of the spin-relaxation model).
In Sec. \ref{S2} we derive a duality relation which maps the phase diagram
onto itself in a non-trivial way. This duality is different from the
domain-wall duality which maps the spin-relaxation dynamics to the BARW.
On a self-dual line running across the
phase diagram we obtain a relation between the (time-dependent) density
expectation value for a half-filled random initial state and the survival
probability of two single particles in an initially otherwise empty
system.

In Sections \ref{S3} and \ref{S4} we adopt the strategy of considering
separately the fluctuations in the total number of particles from
the spatial correlations within a configuration of a given fixed number
of particles. By separating the hopping time scale from the
branching and annihilation time scales one can then gain insight
in the behaviour of the system in the absence of spatial correlations.
In Sec. \ref{S3} we study the system in the fast-diffusion limit
$D \to \infty$ of the BARW (\ref{1-2}) (\ref{S3A}) and in its
spin-relaxation formulation (\ref{1-1}) in the dual limit of infinite
spin-exchange rate $\alpha$.  In these limiting cases all spatial
correlations are washed out and one expects the PC transition to change
into some other, mean-field-type phase transition. However, in contrast
to a traditional mean-field approach, our treatment keeps track of the
exact fluctuations of the {\em total} number of particles (or spins
respectively). Our treatment is not an approximation, but yields
rigorous results in these limiting cases for which
we calculate the stationary density and density fluctuations (\ref{S3A})
and fluctuations of the magnetization (\ref{S3B}). Thus we are able
to analyze to which extent the system deviates from mean-field behaviour
and we also identify the exact phase
transition point. In Sec. \ref{S4} we investigate by similar means the
dynamical behaviour of the finite system in the active region of the
phase diagram. We show that for fast diffusion
the relaxation to the absorbing state in a finite system
is indeed exponentially slow, thus confirming the intuitive argument for
the signature of the active region in a finite system.

In Section \ref{S5} we conclude with some final remarks.

\section{Duality relations}
\label{S2}

We define the BARW in terms of a master equation for the probability
$P(\eta;t)$ of finding, at time $t$, a configuration $\eta$
of particles on a lattice of $L$ sites. Here $\eta = \{\eta(1), \eta(2),
\dots , \eta(L)\}$ where $\eta(x)=0,1$ are the integer-valued particle
occupation numbers at site $x$. For definiteness we assume $L$ to be even.
Using standard techniques \cite{qhf} we express the time
evolution given by the master equation in terms of a Hamiltonian $H$ of a
one-dimensional quantum spin system (for a review see \cite{Schu98}).
The idea is to represent each of the possible particle configurations
$\eta$ by a column vector $\ket{\eta}$ which together with the transposed
vectors $\bra{\eta}$ form an orthonormal basis of a vector space
$X=({\mathbb{C}}^2)^{\otimes L}$. One represents the probability
distribution by a state vector $| \, P(t)\, \rangle = \sum_{\eta \in X}
P(\eta;t) \ket{\eta}$ and writes the master equation in the form
\bel{2-1}
\frac{d}{dt} P(\eta;t) = - \langle \, \eta \, |
H | \, P(t) \, \rangle
\ee
where the off-diagonal matrix elements of $H$ are the (negative)
transition rates between states and the diagonal entries are the inverse
of the exponentially distributed life times of the states.
A state at time $t'=t_0 + t$ is given in terms of an initial
state at time $t_0$ by
\bel{2-2}
| \, P(t_0+t) \, \rangle = \mbox{e}^{-H t }
| \, P(t_0) \, \rangle.
\ee
The expectation value $\rho_k(t)=\bra{s} n_k \ket{P(t)}$ for the density
at site $k$ is given by the projection operator $n_k$ which has value 1
if there is a particle at site $k$ and 0 otherwise. The summation vector
$\bra{s} = \sum_{\eta \in X} \bra{\eta}$ performs the average over all
possible final states of the stochastic time evolution.
Below an initial distribution with $N$ particles placed on sites
$k_1,\dots,k_N$ is denoted by the column vectors $\ket{k_1,\dots,k_N}$.
The empty lattice is represented by the vector $\ket{0}$. The
uncorrelated product distribution where on each lattice site the
probability of finding a particle is equal to 1/2, is given in terms of
the transposed of the summation vector as $\ket{1/2} = \bra{s}^T/(2^L)$.

To obtain the Hamiltonian for the time evolution of the BARW (\ref{1-2})
we note that we can represent any two-state particle system as a spin
system by identifying a particle (vacancy) on site $k$ with a spin-up
(down) state on this site. This allows for a representation of $H$
in terms of Pauli matrices where $n_k =( 1 - \sigma^z_k )/2$
projects on states with a particle on site $k$ and $v_k = 1-n_k$ is the
projector on vacancies. The off-diagonal matrices
$s^{\pm}_k = (\sigma^x_k \pm i \sigma^y_k)/2$
create ($s^-_k$) and annihilate ($s^+_k$) particles. We stress that
in the present context the ``spins'' are just convenient labels for
particle occupancies which are conceptually entirely unrelated to the
spins of the spin relaxation model (\ref{1-1}) which is treated
below. Using this pseudospin formalism one finds
\bea
H & = &  \frac{1}{2} \sum_{k} \left\{D (n_k v_{k+1} +
  v_k n_{k+1}- s^+_ks^-_{k+1}-s^-_ks^+_{k+1})\right. \nonumber \\
\label{2-3}
 &  & \left. + 2\lambda (n_k n_{k+1} - s^+_k s^+_{k+1})
      + \alpha (1-\sigma^x_k\sigma^x_{k+1})n_k\right\}.
\eea
Each part of this stochastic Hamiltonian represents one of the
elementary processes (\ref{1-2}) and we may write
\bel{2-3a}
H(D,\lambda,\alpha) = D H^{SEP} + \lambda H^{RSA} + \alpha H^{BARW}.
\ee
Here $H^{SEP}$ represents hopping of hard-core particles, i.e. the
symmetric exclusion process \cite{Spit70}, the pair-annihilation
process encoded in $H^{RSA}$ corresponds to random-sequential adsorption
\cite{Evan93} and $H^{BARW}$ describes a special equilibrium
branching-annihilating random walk where pair annihilation
requires the presence of another particle and hopping occurs only
in pairs (see (\ref{1-2}).
For $\lambda = D$ and $\alpha=0$ this process reduces to the exactly
solvable process of diffusion-limited pair annihilation (DLPA)
\cite{Torn83}.
The time evolution conserves particle number modulo 2. Here we
work only on the even subspace defined by the
projector $(1+ Q)/2$ where $Q=(-1)^N=\prod_k\sigma_k^z$.
The projection on the even sector of the
uncorrelated initial state with a density $1/2$ is given by the vector
$\ket{1/2}^{even}=(1/2)^{L-1}\ket{s}^{even}$.

Within the same framework the stochastic Hamiltonian for the spin-flip
process (\ref{1-1}) can be written in terms of Pauli matrices as follows
\bea
H^{SF} & = & \frac{1}{4} \sum_{k}\left[ (1-\sigma^x_k) w_k(D,\lambda)
             + \right. \nonumber \\
\label{2-4}
  &   & \left. + \alpha (1-\sigma^x_k\sigma^x_{k+1})(1
                          -\sigma^z_k\sigma^z_{k+1}) \right]
\eea
with the generalized Glauber spin flip rates encoded in
$w_k(D,\lambda) = (2-\sigma^z_{k-1}\sigma^z_{k}-\sigma^z_k\sigma^z_{k+1})
(D+\lambda+(\lambda-D)\sigma^z_{k-1}\sigma^z_{k+1})/2$.
For this process
the spins represent the actual spin configurations of the spin-relaxation
process. We note that the usual zero-temperature Glauber dynamics
- equivalent to DLPA in particle language -
correspond to $\lambda = D$, $\alpha=0$.
For this model the domain wall
correspondence \cite{Racz85} between the BARW and the spin-flip
process can be rigorously derived as a similarity transformation on
the level of the quantum Hamiltonian description. There
exists a transformation ${\cal B}$ such that $H^{SF} = {\cal B} H
{\cal B}^{-1}$ \cite{Sant97}. The generalization
$\lambda = D$, $\alpha > 0$ corresponds to the exactly solvable
process introduced in Ref. \cite{Droz89}.

Consider now the transformation ${\cal D}_{\pm}$
which is, for the even particle sector, defined by
\bel{2-5}
{\cal D}_+ = \gamma_1 \gamma_2 \dots \gamma_{2L-1}
\ee
where
\bea
\gamma_{2k-1} & = & \frac{1}{2} \left[ (1+i) \sigma_k^z
- (1-i) \right] \\
\gamma_{2k} & = & \frac{1}{2} \left[ (1+i) \sigma_k^x\sigma_{k+1}^x
- (1-i) \right].
\eea
and defined by ${\cal D}_-=-{\cal D}_+\sigma_L^x$
for the odd particle sector \cite{Levy91,Schu93}. ${\cal D}_\pm$ is
unitary and transforms Pauli matrices as follows:
\bea
\label{2-6}
{\cal D}_\pm^{-1} \sigma_k^x\sigma_{k+1}^x {\cal D}_\pm & = & \left\{
\ba{ll} \sigma_k^z  & k \neq L \\
        Q\sigma_L^z & k = L
\ea \right. \\
\label{2-7}
{\cal D}_\pm^{-1} \sigma_{k+1}^z {\cal D}_\pm & = & \left\{
\ba{ll} \sigma_k^x\sigma_{k+1}^x  & k \neq L \\
       \pm Q\sigma_L^x\sigma_1^1 & k = L
\ea \right. .
\eea

In Ref. \cite{Schu97} it was observed that this transformation maps
$H^{DLPA}$, obtained from $H$ by setting
$\lambda = D$ and $\alpha=0$, onto its transposed, $H^{DLPA} =
{\cal D} (H^{DLPA})^T {\cal D}^{-1}$ and thus generates a set of
relations between various expectation values \cite{rem1}.
Here we go further and apply this transformation to the
Hamiltonian $H=H(D,\lambda,\alpha)$ (\ref{2-3}) and transpose
the operator which results from the transformation.
Using (\ref{2-6}), (\ref{2-7}) we find
\bea
H^{SEP} & \to & H^{BARW}, \nonumber \\
H^{BARW} & \to  & H^{SEP}, \nonumber \\
H^{RSA} & \to & H^{RSA} + H^{SEP} - H^{BARW} \nonumber
\eea
and hence the relations
\bel{2-8}
\tilde{H} =
\lambda H^{RSA} + (\lambda +\alpha) H^{SEP} +
(D-\lambda) H^{BARW}.
\ee
which has the same form as the original Hamiltonian (\ref{2-3a}), but
with rates
\bea
\tilde{\lambda} & = & \lambda, \nonumber \\
\label{2-8a}
\tilde{D} & = & \lambda+\alpha, \\
\tilde{\alpha} & = & D-\lambda. \nonumber
\eea
The transformation (\ref{2-8}) is a duality
transformation, we obtain the identity transformation if we
apply the transformation twice.

On the mean-field phase transition line $\lambda=D$ \cite{Meny95}
the system is exactly
solvable \cite{Droz89} and belongs in its entirety to the inactive phase.
Hence the whole region $\lambda>D$ is in the inactive phase.
Duality maps the interesting region $\lambda \leq D$
which contains the phase transition line non-trivially onto itself.
Thus duality can be used to relate physical quantities at different
points of the parameter space. This region contains a self-dual line
\bel{2-9}
D = \lambda+\alpha
\ee
in which every point maps onto itself. In the notation of Ref.
\cite{Meny94} $D = p_{rw} = \Gamma^{-1}(1-\delta)$, $\lambda = p_{an} =
\Gamma^{-1}(1+\delta)$ and $\alpha= 2p_{ex} = 2(1-2\Gamma^{-1})$ is
normalized such that $p_{ex}+p_{rw}+p_{an}=1$. With a parametrization in
terms of $\delta$ and $p_{ex}$ the dual rates are given by $\tilde{\delta}
= - 2p_{ex}/[1 + p_{ex} +\delta(1-p_{ex})]$ and $\tilde{p_{ex}} =
- \delta(1-p_{ex})/[\delta(1-p_{ex})+2(1+p_{ex})]$. The self-dual line
is given by the relation $\delta = - 2p_{ex}/(1-p_{ex})$.

The duality transformation not only maps the phase-diagram onto itself,
but also generates relations between time-dependent expectation values.
Consider the expectation value of the density $\rho_k(t) =
\bra{s}^{even} n_k e^{-Ht}\ket{1/2}^{even}$,
where $\ket{1/2}^{even}$ is a random initial state with density 1/2,
projected over the even sector. This expectation value
is defined at the point $(D,\lambda,\alpha)$
of the parameter space of the Hamiltonian. It is straightforward to
verify the relations
\bea
{\cal D}^{-1}\ket{s}^{even} & = & -i(i-1)^{L-1}\ket{0} \nonumber \\
\bra{1/2}^{even}{\cal D} & = &
i(-i-1)^{L-1}\bra{0}/2^{L-1}.
\eea
So if we use these rules
of transformation and the rules of transformation for the Pauli
matrices, given by (\ref{2-6}), (\ref{2-7}),
we can write the expectation value for the density in the even sector
the form
\bea
\bra{s} n_k e^{-Ht} \ket{1/2}^{even} & = &
\frac{1}{2}\bra{0} e^{-\tilde{H}t} (1-\sigma_{k-1}^x\sigma_{k}^x)
\ket{0}  \nonumber \\
\label{2-12}
& = & \frac{1}{2} \left( 1- \bra{0} e^{-\tilde{H}t}
\ket{k,k+1} \right)
\eea
where we have used (\ref{2-8}) and the fact that the expectation value
for the density $\rho_k(t)$ is a real number. The transformed initial
state is a superposition of the steady state (the empty lattice) and the
two-particle state with particles at sites $k,k+1$. The quantity in the
right hand side of equation (\ref{2-12}) is one-half times the
probability that the state with two particles initially placed at sites
$k$ and $k+1$ has not decayed at time $t$ to the empty state,
measured with the transformed rates
$\tilde{\lambda},\tilde{D},\tilde{\alpha}$ (\ref{2-8a}).

This is a specific result for the time-dependent density starting from
a random initial state with density 1/2. More general transformation
properties of time-dependent correlation functions can be obtained
following the strategy of Ref. \cite{Schu97}. We conclude this section
by pointing out that analogous enantiodromy relations can be derived for
a discrete-time version of the process which corresponds to a sublattice
parallel updating scheme rather than the random sequential updating
represented by the stochastic Hamiltonians (\ref{2-3}), (\ref{2-4}).
Such a parallel updating scheme (which we expect to retain all the
universal features of the model) consists of four steps.
In a first updating sweep, update all spins on the even sublattice
in parallel according to the generalized Glauber rules, but with the
rates $\lambda,D$ now taken as actual probabilities. In a second step
one updates the odd sublattice. In a third step one applies a sublattice
parallel pair-updating scheme to implement the Kawasaki spin-exchange
process with probability $\alpha/2$:
In a first round one divides the lattice into even/odd pairs $(2k,2k+1)$
and exchanges spins within each pair with probability $\alpha/2$.
Finally one updates with the odd/even pairs. This completes
a full updating cycle. The stochastic time evolution of this process
may be written in terms of the transfer matrix
\bel{2-13}
T^{SF} = T^{K}_{odd}(\alpha) \;T^K_{even}(\alpha) \;
         T^G_{odd}(\lambda,D)\;T^G_{even}(\lambda,D)
\ee
where
\be
T^G_{even}(\lambda,D) =
\prod_{k=1}^{L/2} \left[1 - (1-\sigma^x_k) w_k(D,\lambda)/4
\right]
\ee
and an analogous expression for $T^K_{odd}$. The spin exchange
transfer matrix $T^K_{odd}T^K_{even}$ is the well-known
transfer matrix of the six-vertex model \cite{KDN} defined on a diagonal
square lattice. One has
\be
T^K_{even} = \prod_{k=1}^{L/2} \left[1 -
\alpha (1-\sigma^x_{2k}\sigma^x_{2k+1})(1
                          -\sigma^z_{2k}\sigma^z_{2k+1})/4\right].
\ee
The transfer matrix for the related BARW model can be obtained by
applying the similarity transformation ${\cal B}$ of Ref. \cite{Sant97}.
One can then derive a duality relation in the way described above.
The transformed process has the same elementary transitions, but with a
different updating sequence.

\section{Phase transition for fast diffusion}
\label{S3}

It is intuitively clear that the PC phase transition in the system
originates in the complicated structure of the density correlations
which are build up by the competing processes of branching and
annihilation. For a better understanding, consider first $\lambda = 0$.
This reduced process includes (besides diffusion) branching $A \to 3A$
and conditional pair annihilation $3A \to A$ which both require the
presence of a surviving particles to take place. As a result, there are
two stationary distributions: the empty lattice, and the random
distribution where each particle configuration is equally likely.
Since there is no transition channel from  the occupied lattice to the
empty lattice, the system is in the active phase.

On the other hand, for $\alpha = 0$ the system is in the absorbing
phase, the only stationary distribution is the empty lattice.
We conclude that the {\em unconstrained}
pair annihilation process $2A \to 0$ with rate $\lambda$
is responsible for the phase transition to take place.

This scenario is captured in a simple mean-field approach. The exact
equations of motion for the expected particle number $\exval{N(t)}$
reads
\be
\frac{d}{dt} \exval{N(t)} = \sum_i \left[ \alpha \exval{n_i(t)}
- 2(\alpha+\lambda) \exval{n_i(t)n_{i+1}(t)} \right].
\ee
Replacing the
correlators by the product of the density $\rho(t)=\exval{n_i(t)}$
yields the mean-field equation
\be
\frac{d}{dt} \exval{N(t)} = \alpha \exval{N(t)} -
\frac{2(\alpha+\lambda)}{L} \exval{N(t)}^2
\ee
with the stationary mean-field solution for the active phase
\bel{rhomf}
\exval{N}^\ast_{mf} = \frac{\alpha}{2(\alpha+\lambda)} L.
\ee
Since each lattice site can take only one particle and therefore
$n_i = 0,1$, one can use $n_i^2=n_i$ to show that the
mean-field fluctuations $\Delta^\ast_{mf} = \exval{N^2}^\ast_{mf} -
\left(\exval{N}^\ast_{mf}\right)^2$ around the mean are given in terms
of the density $\rho^\ast_{mf} = \exval{N}^\ast_{mf}/L$ by
\bel{sigmf}
\Delta^\ast_{mf} =  \rho^\ast_{mf} (1- \rho^\ast_{mf}) L.
\ee
We conclude that the mean-field phase transition point is given by
$\alpha/\lambda = 0$, consistent with the considerations above. By duality
(\ref{2-8a}) we also recover the mean-field phase transition line of Ref.
\cite{Meny95}.

There are several questions that we want to address in this context.
The first is the form of the {\em exact} phase transition line if both
$\alpha$ and $\lambda$ are very small compared to the diffusion rate $D$.
The second question is the {\em nature} of the phase transition in this
limit. If $D \gg \alpha,\lambda$ the spatial correlations build up by the
annihilation/branching process are wiped out very quickly by diffusive
mixing, leaving a transition which we cannot expect to be a PC transition
anymore. Finally, in the next section we study the crossover time scales
on which a large, but finite system reaches the absorbing state.

\subsection{Phase transition in the BARW}
\label{S3A}

To tackle these questions we observe that for fast diffusion the
process simplifies dramatically: In the absence of spatial correlations
the state of the system is fully characterized by the total particle
number $N$. For fixed $N$, each particle configuration occurs
with equal probability $N!(L-N)!/(L)!$ which is just the inverse
of the number of possibilities of placing $N$ particles on a lattice
of $L$ sites. As a result, the dynamics reduce
to a random walk on the integer set
$0,2,4, \dots, 2K, \dots , L$ of total particle number $N=2K$.
Thus we may represent the dynamics as a random walk on a one-dimensional
lattice of $L/2+1$ sites, where the position of the random walker
marks the number of particles of the BARW process and $0$,
representing the empty lattice, is an absorbing point. It remains only
to calculate the hopping rates $r_N$ and $\ell_N$ from site
$N$ to the right ($N+2$) and left ($N-2$) respectively. The state of
the system is then given by the solution of the master equation
\bea
\frac{d}{dt} P_N(t) & = & r_{N-2} P_{N-2}(t) + \ell_{N+2} P_{N+2}(t)
\nonumber \\
\label{3-0}
& & - (r_{N} +\ell_N) P_{N}(t)
\eea
for the probability of finding $N$ particles in the system.
The average particle number is given by
$\exval{N(t)} = \sum_N N P_N(t)$ \cite{rem2}.

By counting the
number of possibilities of finding two vacancies on neighbouring sites
of an occupied site in a random state of $N$ particles one readily
finds
\be\label{3-0a}
r_N = \frac{\alpha }{2}\frac{N (L - N)(L-N-1)}{(L-1)(L-2)}
\ee
as contribution from the branching process with rate $\alpha/2$. An
analogous consideration gives
\be\label{3-0b}
\ell_N = \frac{1}{2}\left[ \frac{\alpha N (N-1)(N-2)}{(L-1)(L-2)}
+\frac{2\lambda N (N-1)}{(L-1)}\right]
\ee
as the contribution from the annihilation processes $3A \to A$ and
$2A \to 0$ respectively.

These rates represent a biased random walk which in the thermodynamic
limit $L \to \infty$ and for $N$ fixed reduces to a directed random walk
in positive direction with increasing hopping rate $r_N=\alpha N/2$.
Since for $D \to \infty$ the branching process is not diffusion-limited,
the particle number increases exponentially in time,
$\exval{N(t)} = \exval{N(0)} e^{\alpha t}$. Thus for any $\alpha > 0$ the
system is in the active phase, i.e.,
the phase transition is at $\alpha = 0$ which is
consistent with the mean-field result (\ref{rhomf}).
For a large, but finite system
with a small initial number of particles
one expects a slowing down of the exponential growth when a finite
density is reached, i.e. on a time scale of the order $\ln{(L)} /\alpha$.
Ultimately, though, the finite system will reach, by a rare fluctuation
in the number of particles, the absorbing empty lattice.
This second crossover time to absorption  is discussed
in the next section.

To study the stationary behaviour $d/(dt) P_N(t)=0$ of the system we
rescale the lattice to unit length and expand
the r.h.s. of the master equation (\ref{3-0}) in a Taylor series in the
lattice spacing $1/L$. Setting $x=N/L$ and keeping the leading order term
yields the equation $c = 2 ( l_x - r_x ) P^\ast_x$.
The integrability condition $\int_0^1dxP^\ast_x=1$ on the stationary
probability distribution requires the integration constant $c$ to
vanish. The resulting equation
has the solution $P^\ast_x = \delta(x)$, corresponding
to the absorbing state. The only other integrable solution on the interval
$[0,1]$ is the delta-function $P^\ast_x = \delta(x-\rho^\ast)$ with
\bel{3-0d}
\rho^\ast = \frac{\alpha}{2(\alpha+\lambda)}.
\ee
This gives the exact stationary density $\rho^\ast$ of the active phase
which, not very surprisingly, coincides with the mean-field value
(\ref{rhomf}).

To determine whether the system in the infinite-diffusion limit actually
{\em is} a mean-field system we investigate the fluctuations around the
mean (\ref{3-0d}). The mean field result (\ref{sigmf}) requires studying
the fluctuations on a length scale of order $y = \sqrt{L}(x - \rho^\ast)$.
Keeping in the Taylor expansion of the master equation around
$x=\rho^\ast$ all terms to this order gives the ordinary differential
equation
\be
\frac{d}{dy} P^\ast_y = - \frac{y}{2\rho^\ast(1-\rho^\ast)^2} P^\ast_y
\ee
The solution of this equation is a Gaussian which gives the exact
fluctuations in the particle number
\bel{3-0e}
\Delta^\ast = 2\rho^\ast(1-\rho^\ast)^2 L
\ee
Except for $\lambda=0$ ($\rho^\ast=1/2$) this expression is in
disagreement with the
mean-field result (\ref{sigmf}), indicating a non-trivial effect of the
unconstrained pair-annihilation process even in the fast diffusion limit.
We conclude that the system undergoes a mean-field transition, but with
fluctuations in the particle number which deviate from those predicted
by mean-field.

\subsection{Spin-relaxation formulation}
\label{S3B}

We consider the system in the dual limit $\alpha \to \infty$ where we can
study the phase transition between the active phase and the absorbing
phase in terms of the dimensionless variable $u=(D-\lambda)/(D+\lambda)$.
In the spin-relaxation picture this is the limit of fast Kawasaki
spin-exchange where the system is spatially uncorrelated and hence
completely characterized by the total magnetization $M =\sum_k
\sigma_k^z/2$. The dynamics of the process
(\ref{1-1}) reduce in this limit to a random walk in the magnetization
variable $M$, ranging from $-L/2$ to $L/2$. The master equation reads
\bea
\frac{d}{dt} P_M(t) & = & r_{M} P_{M-1}(t) + \ell_{M+1} P_{M+1}(t)
\nonumber \\
\label{3-1}
 & & - (r_{M} +\ell_M) P_{M}(t).
\eea
The transition rates for this random walk with absorbing boundaries
at $M= \pm L/2$ are readily calculated as
\bea
r_M & = & \frac{D+\lambda}{2L-2}\left(1-\frac{2 M u}{L-2}\right)
\left(\frac{L^2}{4}-M^2\right) \\
\ell_M & = & \frac{D+\lambda}{2L-2}\left(1+\frac{2 M u}{L-2}\right)
\left(\frac{L^2}{4}-M^2\right).
\eea
We find a bias towards to the boundaries $M=\pm L/2$, i.e.,
the fully magnetized
absorbing states with all spins up or down resp., for $u < 0$.
For $u > 0$ the system is biased to the center, corresponding to
the active phase.

For an initial state which is symmetric under spin-flip
$s_i \to -s_i$ the mean $\exval{M}$ vanishes for all times in both the
active and absorbing phase and hence is not suitable to characterize the
system. For the same reason an naive mean-field approach
by setting $\exval{\sigma_i^z(t) \sigma_j^z(t)} = \exval{\sigma_i^z(t)}
\exval{\sigma_j^z(t)}$ would not give any information on the dynamics
of the spin-fluctuations. The quantity that characterizes the phase
transition are the fluctuations $\exval{M^2}=\sum_M M^2 P_M(t)$ in the
magnetization, i.e. the mean-square displacement of the random walk.
In the active regime this quantity is proportional to
the system size whereas in an ordered state $\exval{M^2} \sim L^2$.

First consider the phase transition point $u=0$. From the considerations
above we know that the stationary state is inactive. The only question
of interest is the approach to stationarity from some random initial
state. From (\ref{3-1}) one obtains $d/(dt)\exval{M^2} = 2 \lambda(L^2/4
- \exval{M^2})/(L-1)$ which is readily solved by
\be
\exval{M^2(t)} = L^2/4 + (\exval{M^2(0)} - L^2/4)e^{-2\lambda t/(L-1)}.
\ee
The approach to the stationary value is exponential on a time scale
\bel{3-1a}
\tau = \frac{L-1}{2\lambda}.
\ee
For large system size and initial times $\tau \ll L$, the fluctuations
in the magnetization grow linearly in time.

For $u \neq 0$ the equations of motion for the moments $\exval{M^{2k}}$
are too complicated for direct analysis.
We define $\hat{M} = M/\sqrt{L}$ and study only the thermodynamic limit.
Using the master equation (\ref{3-1})
the stationarity condition $d/(dt) \exval{\hat{M}^{2k}} = 0$ for the
moments of $\hat{M}$ yields the recursion relation
\be
\exval{\hat{M}^{2k}} = (2k-1) \exval{\hat{M}^{2k-2}} / (4u)
\ee
which shows that the stationary distribution in the active phase
$u > 0$ is Gaussian with variance $1/(4u)$:
\bel{3-2}
P^\ast(\hat{M}) = \sqrt{\frac{2u}{\pi}} e^{-2u \hat{M}^2}.
\ee
This yields the final result
\be
\exval{\hat{M}^{2}} = \left\{ \ba{cc} 1/(4u) & u > 0 \\
                              \infty & u \leq 0
                      \ea \right.
\ee
All other stationary moments in the active regime follow from the
Gaussian nature (\ref{3-2}) of the statistics.
We read off a critical exponent $\kappa = 1$
for the divergence of $\exval{\hat{M}^{2}}$ with $u$ as the system
approaches the critical point $u=0$.

\section{Relaxational behaviour in finite systems}
\label{S4}

The exact solution \cite{Droz89} for the dynamics of the spin-spin
correlation function on the line $\lambda = D$ implies a crossover
time $\tau \sim L^2$ from a power law relaxation to exponential
relaxation to the absorbing state. One then expects this to hold
throughout the inactive phase.

On the other hand, any finite system has only one stationary state,
which is the empty lattice in particle language, corresponding to the
magnetically
ordered states with all spins up or all spins down. It is therefore of
interest to study the relaxation towards this state in that region
of parameter space that constitutes the active phase of the infinite
system.

The precise location of the phase transition line is not known, but
we know that the line $\lambda = 0$ (no unrestricted pair
annihilation) belongs to the active phase and thus we may get some
insight by studying the system in the immediate neighbourhood $\lambda
\ll \alpha, D$ of this line. To this end we adopt a similar strategy
as in the previous section by assuming $\lambda$ to be so small that
the system had sufficient time to relax to its $\lambda = 0$ stationary
distribution between two successive pair-annihilation events.
This limiting procedure can be made rigorous by taking $\alpha \to
\infty$ and keeping $\lambda,D$ fixed.

In this limit the system reduces effectively to a two-state system,
i.e. the system is completely characterized by stating whether the
system is empty (state $\ket{0}$) or not (which we denote by
$\ket{1}$). The latter state represents the stationary distribution of
the system with $\lambda=0$ in which, because of detailed balance for
this reduced process, all states with an even, non-zero
number of particles have equal probability $p=1/[2^{(L-1)} - 1]$.
The transition rates between these two states characterizing the system
are then trivial to work out: The transition from $\ket{0}$ to $\ket{1}$
is zero because $\ket{0}$ is an absorbing state. On the other hand,
counting the number of states represented by $\ket{1}$ for which a pair
annihilation event leads to the empty lattice yields a transition
rate
\bel{4-1}
1/\tau_{act} = (\lambda L) / [2^{(L-1)} - 1]
\ee
for transitions from state 1
to state 0. Hence, at time $t$ the system is in the
absorbing state with probability $P_0(t) = 1 - e^{-t/\tau_{act}}$
and in each non-empty state with probability $P_1(t) =
e^{-t/\tau_{act}}/ [2^{(L-1)} - 1]$.

For the particle density this behaviour implies the exact result
\bel{4-2}
\rho(t) = \frac{ e^{-t/\tau_{act}}}{2-(1/2)^{L}}.
\ee
Because of the fast intermediate relaxation to the equilibrium state of
the $\alpha=\infty$ process the density and the density-correlations
have no spatial dependence and the diffusion rate does not enter.
The crossover time $\tau_{act}$ for
reaching the absorbing state in the active region of the phase diagram
for the infinite system (small $\lambda$) diverges exponentially in
system size.

Finally, we study the dynamical behaviour of the system for large, but
finite $L$ in the limit of fast diffusion dicussed in Sec. \ref{S3A}.
We recall relation (\ref{2-12}) which relates the decay of
the particle density to the survival probability of two neighouring
particles in an empty lattice. This quantity can be interpreted as a
first-passage-time distribution for two annihilating and branching random
walkers: When two random walkers in an empty lattice annihilate for the
first time, the dynamics stop. Therefore the density decay equals one half
this first-passage-time distribution and the mean-first-passage-time
(MFPT)
\bea
\tau & = & \int_0^\infty dt \bra{0} e^{-Ht} \ket{k,k+1}\nonumber \\
\label{4-6}
       & = & \lim_{c->0} \bra{0} (H+c)^{-1} \ket{k,k+1}
\eea
gives the crossover time scale on which the system reaches the absorbing
state.

This quantity can be evaluated numerically for any point in
parameter space by inverting the
time evolution operator $c+H$ for finite system size, then taking the
matrix element (\ref{4-6}) and finally calculating the limit $c \to 0$.
For an analytical treatment for large $D$ we note that the MFPT from
some site $k$ to an absorbing site $k=0$ for a general random walk with
nearest neighbour hops on $L+1$ sites can be expressed in terms of the
hopping rates \cite{MFPT}. In the mapping of the BARW to the random walk
the MFPT $\tau$ is equal to the MFPT of the random walker starting at
site 2. With the hopping rates (\ref{3-0a}), (\ref{3-0b})
we find after some rearrangement of terms
\bel{4-3}
\tau = \frac{1}{\lambda L} \sum_{k=0}^{L/2-1} c_k
\ee
with
\bel{4-4}
c_k = \frac{L! \,k! \, \Gamma(\lambda(L-2)/\alpha+1)}
{(2k+2)!\,(L-2k-2)! \, \Gamma(\lambda(L-2)/\alpha+k+1)}.
\ee

We note first that in the limit $\alpha \to \infty$ the MFPT coincides
with the relaxation time (\ref{4-1}), as expected from duality.
To study the asymptotic behaviour of $\tau$ for finite $\alpha$ (active
phase) we determine the value $k_{0}$ for which $c_k$ gives the largest
contribution to sum on the r.h.s. of (\ref{4-3}). We find
$k_{0} = \rho^\ast L/2$ with the stationary density given by (\ref{3-0d}).
Using the Stirling formula for the Gamma-function and expanding $c_k$
around $k_{0}$ yields for non-vanishing density $\rho^\ast$ the asymptotic
form of the crossover time
\bel{4-5}
\tau_{act} \sim \frac{1}{\lambda}
\left[\frac{(1-2\rho^\ast)^{(1-2\rho^\ast)/(2\rho^\ast)}}
{(1-\rho^\ast)^{(1-\rho^\ast)/(\rho^\ast)}}\right]^L
\ee
up to subleading power-law corrections in system size.
Therefore, in the active region of the phase diagram the
crossover to absorption in a finite system takes place on a time scale
which is exponentially large in system size, with a density-dependent
amplitude.

For $\rho^\ast = 0$, i.e. in the absorbing phase, the MFPT can be
read off directly from (\ref{4-3}), (\ref{4-4}), since only the
term with $k=0$ contributes. One finds
\be
\tau_{abs} = \frac{(L-1)}{2\lambda}
\ee
This power law differs from the crossover behaviour $\tau_{abs}
\sim L^2/D$ for finite diffusion constant D. The MFPT
for this point in parameter space coincides with the relaxation time
(\ref{3-1a}) in the dual point.

\section{Final remarks}
\label{S5}

The duality relation (\ref{2-8}) divides the parameter space
into two distinct regions separated by the self-dual line (\ref{2-9}).
Both regions are mapped onto each other and hence have the same
physical properties. This is of practical usefulness for a numerical
survey of the system and for the determination of the phase transition
line since only part of the parameter space needs to be investigated.

In particular, the line $\alpha = 0$ maps onto the
line $D=\lambda$ and the fast-diffusion limit to
the limit $\alpha \to \infty$. The observation that for large $D$
any small $\alpha$ brings the system into the active phase translates
into a phase transition at $D=\lambda$ in the limit $\alpha \to \infty$.
This is a new result which definitely clarifies the unresolved issue of
the location of the phase transition line for large $\alpha$.
Numerical analysis of the model for large rates $\alpha$ is reported to
be very difficult \cite{Meny94,Meny95}. Our exact result confirms the
conjecture of Ref. \cite{Meny94} on the location of the phase transition
point in this numerically untractable limit. In the $\delta\;-\;p_{ex}$
phase diagram of Ref. \cite{Meny94} the limit $\alpha \to \infty$
corresponds to $p_{ex} \to 1$ and the phase transition point $D=\lambda$
corresponds to $\delta = 0$. The dual limit $D \to \infty$ covers the
neighbourhood of the point
$\delta = -1, p_{ex}=0$. Our result translates into an infinite slope
of the phase transition line in this representation at this point.
In the active phase of this region the exact stationary
particle number distribution is Gaussian with a stationary density
$\rho^\ast$ given by the mean-field value (\ref{rhomf}). The
density fluctuations (\ref{3-0e}) deviate from the mean-field result
(\ref{sigmf}) by a factor $2(1-\rho^\ast)$.

On the self-dual line we find from (\ref{2-12}) that the density
expectation value $\rho_k(t)$ equals one-half the survival probability at
time $t$ of two particles placed initially at two neighbouring sites
in an otherwise empty lattice. Hence the phase transition from the
absorbing phase to the active phase may be rephrased as a
mean-first-passage-time (MFPT) transition for random walkers which branch
and annihilate. We expect the MFPT in a finite system to
change from a power-law divergence (in system size) to an exponential
divergence not only at infinite diffusion rate (Sec. \ref{S4}), but also
for finite $D$. Thus this numerically accessible quantity provides an
alternative way of determining and characterizing the PC phase transition.

\section{Acknowledgements}

We thank J. Cardy, N. Menyh\'ard, Z. R\'acz and U. T\"auber for useful
discussions. G.M.S. thanks the Institute for Theoretical
Physics of the E\"otv\"os University for hospitality and
partial support (Grant OTKA OTKA T 019451).



\begin{references}

\bibitem{DP} T. E. Harris, Ann. Prob. {\bf 2}, 969 (1974);
H. Takayasu and A. Yu. Tretyakov {\bf 68},
3060 (1992); I. Jensen, Phys. Rev E {\bf 47}, R1 (1993), Phys. Rev. Lett.
{\bf 70}, 1465 (1993).

\bibitem{PC} P. Grassberger, F. Krause and T. von der Twer,
J. Phys. A {\bf 17}, L105 (1984); P. Grassberger, J. Phys. A {\bf 22},
L1103 (1989); M. H. Kim and H. Park, Phys. Rev. Lett. {\bf 73},
2579 (1994); I. Jensen, J. Phys. A {\bf 26}, 3921 (1993),
Phys. Rev. E {\bf 50}, 3623 (1994); D. ben-Avraham, F. Leyvraz and
S. Redner, Phys. Rev. E {\bf 50}, 1843 (1994). Some of the references
consider BARW's with both even and odd number of offsprings.

\bibitem{Meny94}
N. Menyh\'ard, J. Phys. A {\bf 27}, 6139 (1994).

\bibitem{Card97}
J. Cardy and U. T\"auber, Phys. Rev. Lett. {\bf 77}, 4780 (1996);
J. Stat. Phys. {\bf 90} (1998) (in press).

\bibitem{Glau63}
R. J. Glauber, Math. Phys. {\bf 4}, 294 (1963).

\bibitem{Kawa66}
K. Kawasaki, Phys. Rev. {\bf 145}, 224 (1966).

\bibitem{Racz85}
Z. R\'acz, Phys. Rev. Lett. {\bf 55}, 1707 (1985).

\bibitem{Droz89}
M. Droz, Z. R\'acz and J. Schmidt, Phys. Rev. A {\bf 39}, 2141 (1989).

\bibitem{Lee94}
B. P. Lee, J. Phys. A {\bf 27}, 2633 (1994).

\bibitem{qhf} E. Siggia, Phys. Rev. B {\bf 16}, 2319 (1977);
S. Sandow and S. Trimper, Europhys. Lett. {\bf 21}, 799 (1993);
F. C. Alcaraz, M. Droz, M. Henkel and V. Rittenberg,
Ann. Phys. (N.Y.) {\bf 230}, 250 (1994);
G. M. Sch\"utz, J. Stat. Phys. {\bf 79}, 243 (1995).

\bibitem{Schu98}
G. M. Sch\"utz, {\it Integrable Reaction-Diffusion
Processes and Quantum Spin Chains} (to be published).

\bibitem{Spit70}
F. Spitzer, Adv. Math. {\bf 5}, 246 (1970).

\bibitem{Evan93} J. W. Evans, in {\it Nonequilibrium Statistical
Mechanics in One Dimension}, ed. V. Privman
(Cambrige University Press, Cambridge UK, 1996).

\bibitem{Torn83} D. C. Torney and H. M. McConnell,
 J. Phys. Chem. {\bf 87}, 1941 (1983).

\bibitem{Sant97}
J. E. Santos J. Phys. A {\bf 30}, 3249 (1997).

\bibitem{Levy91}
D. Levy, Phys. Rev. Lett. {\bf 67}, 1971 (1991).

\bibitem{Schu93}
G. Sch\"{u}tz J. Phys. A {\bf 26}, 4555 (1993).

\bibitem{Schu97}
G. M. Sch\"utz, Z. Phys. B {\bf 104}, 583 (1997); G. M. Sch\"utz and
K. Mussawisade, Phys. Rev. E {\bf 57} (1998) (in press).

\bibitem{rem1} Such a relationship between a stochastic processes is
called an enantiodromy relation, as opposed to a similarity transformation
like the domain-wall duality which relates one stochastic Hamiltonian
to another, rather than to its transposed.

\bibitem{Meny95}
N. Menyh\'ard and G. \'Odor J. Phys. A {\bf 28}, 4505 (1995).

\bibitem{KDN}
D. Kandel, E. Domany and B. Nienhuis, J. Phys. A {\bf 23}, L755 (1990).

\bibitem{rem2} This and the other limits of fast rates discussed in this
paper can be treated rigorously by taking the limit of large rates
in the formal solution \protect\ref{2-2} of the master equation,
see \cite{Schu98} for details.

\bibitem{MFPT}
G. H. Weiss, J. Stat. Phys. {\bf 24}, 587 (1981); K. P. N. Murthy and K. W.
Kehr, Phys. Rev. A {\bf 40}, 2082 (1989), and erratum in Phys. Rev. A {\bf 41},
1160 (1990).

\end{references}
\end{document}